\documentclass[twocolumn, aps, PRL, floatfix]{revtex4-2}
\usepackage{amsmath}
\usepackage{amssymb}
\usepackage{lipsum}
\usepackage{graphicx}
\usepackage{dcolumn}
\usepackage{bm}
\usepackage{xcolor}
\usepackage{epstopdf}
\usepackage{ulem}
\begin{document}

\preprint{Phys.Rev.Lett.}

\title{Resistivity of non-Galilean invariant two dimensional Dirac system}

\author{V. M. Kovalev,$^{1,2}$ M. V. Entin$^{1,3}$, Z. D. Kvon,$^{1,3}$  A. D. Levin,$^4$ V. A. Chitta,$^4$ G. M. Gusev,$^4$  and  N. N. Mikhailov$^{1,3}$}

\affiliation{$^{1}$Institute of Semiconductor Physics, Novosibirsk
630090, Russia}
\affiliation{$^2$Novosibirsk State Technical University, Novosibirsk 630073,
Russia}
\affiliation{$^3$Novosibirsk State University, Novosibirsk 630090,
Russia}
\affiliation{$^4$Instituto de F\'{\i}sica da Universidade de S\~ao
Paulo, 135960-170, S\~ao Paulo, SP, Brazil}

\date{\today}
\begin{abstract}
We revisited the influence of electron-electron scattering on the resistivity of a two-dimensional system with a linear spectrum. In conventional systems with a parabolic spectrum, where Umklapp scattering is either prohibited or ineffective due to the small Fermi surface, particle-particle scattering does not contribute to conductivity because it does not change the total momentum. However, within the framework of the Boltzmann kinetic model, we demonstrate that electron-electron scattering in Dirac systems can significantly contribute to conductivity, producing distinct temperature-dependent corrections: a $T^4$ behavior at low temperatures and a $T^2$ dependence at moderate temperatures. While the predicted $T^4$ scaling is not observed experimentally—likely suppressed by dominant weak localization effects—the $T^2$ scaling is clearly confirmed in our measurements. Specifically, temperature-dependent resistivity data from a gapless single-valley HgTe quantum well exhibit $T^2$ corrections, which align well with theoretical predictions. Thus, we challenge the paradigm that the $T^2$ term in resistivity is absent in single-band 2D metals.
\end{abstract}

\maketitle

\textit{Introduction.---} The transport properties of strongly interacting systems pose a significant challenge in modern physics. One long-standing issue is the century-old debate regarding how electron-electron scattering contributes to resistivity, which is often theorized to follow a $\rho \sim T^2$ behaviour \cite{haas}. Despite this expectation, experimentally detecting this $T^2$ behavior and attributing it directly to electron-electron interactions has proven difficult and remains largely inconclusive \cite{pinski, behnia}.

This quadratic temperature resistivity contribution only emerges when the Fermi surface is open and crosses the boundary of the Brillouin zone. Under these conditions, electrons can transfer momentum to the lattice by jumping between opposite sections of the Fermi surface. However, the resulting resistance is not a universal property; it is highly sensitive to the details of the Fermi surface topology. Furthermore, a collision integral representing electron-electron interactions that do not transfer momentum to the lattice fails to impact resistivity, unless the electric current is independent of the momentum density \cite{pal}.

Consequently, the influence of electron-electron interactions on transport properties can be identified through several mechanisms for non Galilean invariant systems: (a) the presence of specific Fermi surface characteristics that enable Umklapp scattering \cite{pal}; (b) in compensated semimetals \cite{olshanetsky, entin}; (c) in two-subband systems with different effective masses and  impurity scattering time \cite{hwang, nagaev1, nagaev2, gusev, levin},  and (d) through electronic hydrodynamic effects in narrow channels\cite{gurzhi, polini, narozhny, dejong, kumar, spivak, gusev1, gusev2, gusev3, gusev4} .

It is important to note, that perturbation theory predicts that, unlike
the Galilean-invariant case where not only the $T^2$ term but all
higher-order terms are absent, the finite-T terms may appear for a non-parabolic spectrum.
A qualitative theoretical analysis in \cite{pal} states that it should be a $T^4$ behavior.
However, this situation has not yet been studied in details. Here, we prove both
theoretically (using the same approach as in paper \cite{pal})
and experimentally that the resistivity of degenerate electron system
with linear-in-momentum spectrum behaves as $T^2$.

Materials with a Dirac-like spectrum, such as graphene, require more detailed theoretical and experimental study due to several specific features in their spectrum and transport properties. Firstly, compensated graphene is a pure material and should exhibit resistance dominated by electron-electron interactions. It is essential to recognize that at the charge neutrality point, thermal excitation of both electrons and holes facilitates the investigation of electron-electron limited transport in the non-generated regime \cite{bolotin}.  Furthermore, graphene samples are fabricated using exfoliation techniques, resulting in mesoscopic dimensions. In such cases, the channel size affects electron-electron limited transport (Gurzhi effect), while a thorough transport analysis generally necessitates macroscopic samples. Additionally, much of the research has concentrated on bilayer graphene, where the energy spectrum approaches parabolic characteristics at low energies \cite{wagner, nam, tan}.

Recognizing the importance of new quantum materials with a linear spectrum, it is essential to search for materials that can serve as effective platforms for studying interaction-affected transport. Gapless HgTe quantum wells present such an opportunity due to their high mobility and the ability to control carrier density. In this paper, we theoretically investigate the contribution of electron-electron scattering in systems with a linear spectrum far from the Dirac point. We derive a $T^2$ term in the conductivity, a conclusion that is quite general and applicable to any system exhibiting a linear spectrum. Experimentally, we tested this prediction in a specific gapless HgTe quantum well and found reasonable agreement with our theoretical results. Thus, our findings confirm that the $T^2$ term in resistivity is indeed present in single-band 2D Dirac metals.

\textit{Theory.---} A theoretical model we consider is as follows. We assume that there is a two-dimensional single band material with a linear dispersion characterized by the constant band parameter $v$ having the velocity dimension. Thus, in framework of this model, the energy dispersion of degenerate charge carriers with momentum $\bf p$ is given by the expression  $\varepsilon_{\bf p}=vp$. We assume the charge carriers to be electrons. Such a theoretical model is directly related to the experimental structure considered in the next section.

At zero temperature, the system resistance is determined by the electron scattering off impurities characterized by the constant scattering time, $\tau_i$. At finite temperatures the temperature-dependent correction to the resistance is due to inter-particle scattering processes. At finite but low enough temperatures, when the electron-electron scattering processes is characterized by the scattering time $\tau_{ee}$, the temperature-dependent correction to the resistivity can be found via a successive approximation with respect to small parameter $\tau_{ee}\gg\tau_i$, considering the inter-electron scattering integral in Boltzmann transport equation perturbativly.

The Bolzmann equation, describing the electron interaction with external electric field ${\bf E}$, random impurity potential and with other electrons, reads $-({\bf F}\cdot\nabla_{\bf p})f_{\bf p}=-(f_{{\bf p}}-n_{{\bf p}})/\tau_i+Q_{ee}\{f_{\bf p}\}$.

Here ${\bf F}=e{\bf E}$, where $e>0$ is an absolute value of electron charge, $n_{{\bf p}}$ is an equilibrium electron distribution function and $Q_{ee}$ is a electron-electron collision integral. We apply here the simple $\tau_i$-approximation for the electron-impurity collision integral. 

In the limit $\tau_{ee}\gg\tau_i$, the interparticle collision integral can be considered as a correction. Expanding nonequilibrium distribution function into the first order with respect to the external force ${\bf F}$ as $f_{\textbf{p}(\textbf{k})}=n_{\textbf{p}}+\delta f_{\textbf{p}}$, the linear-in-${\bf F}$ correction is decomposed into the electron-impurity scattering contribution and the contribution related to electron-electron scattering, $\delta f_{\textbf{p}}=\delta f^{(0)}_{\textbf{p}}
+\delta f^{(1)}_{\textbf{p}}$. Here $\delta f^{(0)}_{\textbf{p}}$ and $\delta f^{(1)}_{\textbf{p}}$  are the iteration corrections with respect to small term $Q_{ee}$. Deploying
the method of successive iterations results in $\delta f^{(0)}_{\bf p}=
\tau_i({\bf F}\cdot{\bf v}_{\bf p})n'_{\bf p}$ and $\delta f^{(1)}_{\textbf{p}}=\tau_iQ_{ee}\{\delta f^{(0)}_{\bf p}\}$.

Electron-electron collision induced electric current density correction reads as ${\bf j}=-e\tau_i\sum_{\bf p}{\bf v}_{\bf p} Q_{ee}\{\delta f^{(0)}_{\bf p}\bigr)$. 
After cumbersome calculations \textcolor{red}{\cite{suppl}}, we arrive at the expression describing the current density correction induced by the inter-particle collisions in the form

\begin{gather}\label{current_general_zero}
{\bf j}=2\pi e\tau_i^2\sum_{{\bf p},{\bf k},{\bf q}}
|U_{\mathbf{q}}|^2
(n_\mathbf{p}-n_{\mathbf{p}-\mathbf{q}})
(n_\mathbf{k}-n_{\mathbf{k}+\mathbf{q}})\\\nonumber
\times
({\bf F}\cdot{\bf v}_{\bf p})
[{\bf v}_{\bf p}
+{\bf v}_{\bf k}
-{\bf v}_{{\bf p}-{\bf q}}
-{\bf v}_{{\bf k}+{\bf q}}]\times\\\nonumber
\int\limits_{-\infty}^{+\infty}\frac{d\omega}{4T\sinh^2\left(\frac{\omega}{2T}\right)}
\delta(\epsilon_{\mathbf{k}+\mathbf{q}}-\epsilon_\mathbf{k}-\omega)
\delta(\epsilon_{\mathbf{p}-\mathbf{q}}-\epsilon_\mathbf{p}+\omega).
\end{gather}

Here $U_{\bf q}$ is an Fourier-transform of interparticle interacting potential. This is a general expression for the current correction. It is easily to see that for Galilean-invariant systems, ${\bf v}_{\bf p}={\bf p}/m$ and total velocity under collision in Eq.\eqref{current_general_zero} vanishes ${\bf v}_{\bf p} +{\bf v}_{\bf k}-{\bf v}_{{\bf p}-{\bf q}}-{\bf v}_{{\bf k}+{\bf q}}=({\bf p}+{\bf k}-{\bf p}+{\bf q}-{\bf k}-{\bf q})/m=0$. This is not the case for Galilean-non-invariant system, say for systems with linear dispersion, $\varepsilon_{\bf p}=vp$. Indeed, the corresponding velocity reads ${\bf v}_{\bf p}=v{\bf p}/p=v^2{\bf p}/\epsilon_{\bf p}$, where $p=|{\bf p}|$ is an absolute value of particle momentum. Thus, the velocity factor 
${\bf v}_{\bf p} +{\bf v}_{\bf k}-{\bf v}_{{\bf p}-{\bf q}}-{\bf v}_{{\bf k}+{\bf q}}=
v^2({\bf p}\epsilon^{-1}_{\bf p}+{\bf k}\epsilon^{-1}_{\bf k}-({\bf p}-{\bf q})\epsilon^{-1}_{{\bf p}-{\bf q}}-({\bf k}+{\bf q})\epsilon^{-1}_{{\bf k}+{\bf q}})$
does not vanish under the momentum conservation condition, resulting to the nonzero contribution to the current Eq.\eqref{current_general_zero}. The conductivity correction induced by interparticle scattering, is directly related to the current density Eq.\eqref{current_general_zero} in non-Galilean invariant system. The lengthy analysis (\cite{suppl}) yields ($\varepsilon_q=vq$)

\begin{gather}\label{theorycond}
\sigma_{xx}=-\frac{1}{\pi^3}\left(\frac{e\tau_i}{4 v}\right)^2\left(\frac{1}{T}\right)
\int\limits_{-\infty}^{+\infty}\frac{\omega^4d\omega}{\sinh^2(\omega/2T)}
\int\limits_{|\omega|/v}^\infty\frac{qdq}{2\pi}
\frac{|U_{\bf q}|^2}{\varepsilon^2_q}.
\end{gather}

The final integration requires the particular form of inter-particle interaction potential, $U_{\bf q}=2\pi e^2/\varepsilon (q+q_s)$, where $q_s$ is wavevector related to screening effects due to the presence of mobile carriers. Below we present the comparison of this expression with experimental data. It should be noted that the expression Eq.\eqref{theorycond} was formally found via perturbation procedure over small parameter $\tau_i/\tau_{ee}\ll1$ and may formally holds up to $\tau_i/\tau_{ee}\leqslant 1$. In the region $\tau_i/\tau_{ee}\geqslant 1$ our approach is not applicable and more sophisticated self-consisted method must be used \cite{Fink1, Fink2}.
Taking into account the bare Drude conductivity for particles with linear Dirac spectrum
$\sigma_0=\frac{e^2}{4\hbar}\left(\frac{\mu\tau_i}{\hbar}\right)$, where $\mu$ is Fermi level of degenerate electrons, the correction to resistivity reads:
\begin{gather}\label{int}
\frac{\delta \rho}{\rho_0} = \frac{8}{\pi^2}\left(\frac{e^2}{\epsilon \hbar v} \right)^2 \left(\frac{T}{\mu}\right) \left( \frac{T\tau}{\hbar} \right) \mathfrak{J}_T
\end{gather}
\begin{gather}
\mathfrak{J}_T =\left( \frac{T}{T^*} \right)^2 \int_{0}^{\infty} \frac{x^4 \, dx}{\sinh^2(x)} \left[ \ln \left( 1 + \frac{T^*}{xT} \right) - \frac{1}{1 + xT / T^*} \right]
\end{gather}
where $T^{\star} =  \hbar v q_s =\frac{\mu e^2}{\varepsilon \hbar v}$ is the characteristic temperature separating different temperature regimes \cite{suppl}. Parameter $q_s = 2\mu e^2/\epsilon v^2$ represents the screening wavevector for mobile carriers with a linear spectrum. One can expect that at low temperatures ($T \ll T^{\star}$), $\mathfrak{J}_T \rightarrow \left(\frac{T}{T^{\star}}\right)^2\ln T$, and the corrections are given by $\sigma_{xx} \sim T^4 \ln(2\mu/T)$. At high temperatures ($T \gg T^{\star}$), $\mathfrak{J}_T \rightarrow \text{const}$, and the corrections follow $\sigma_{xx} \sim T^2$.

\begin{figure}
\includegraphics[width=9cm]{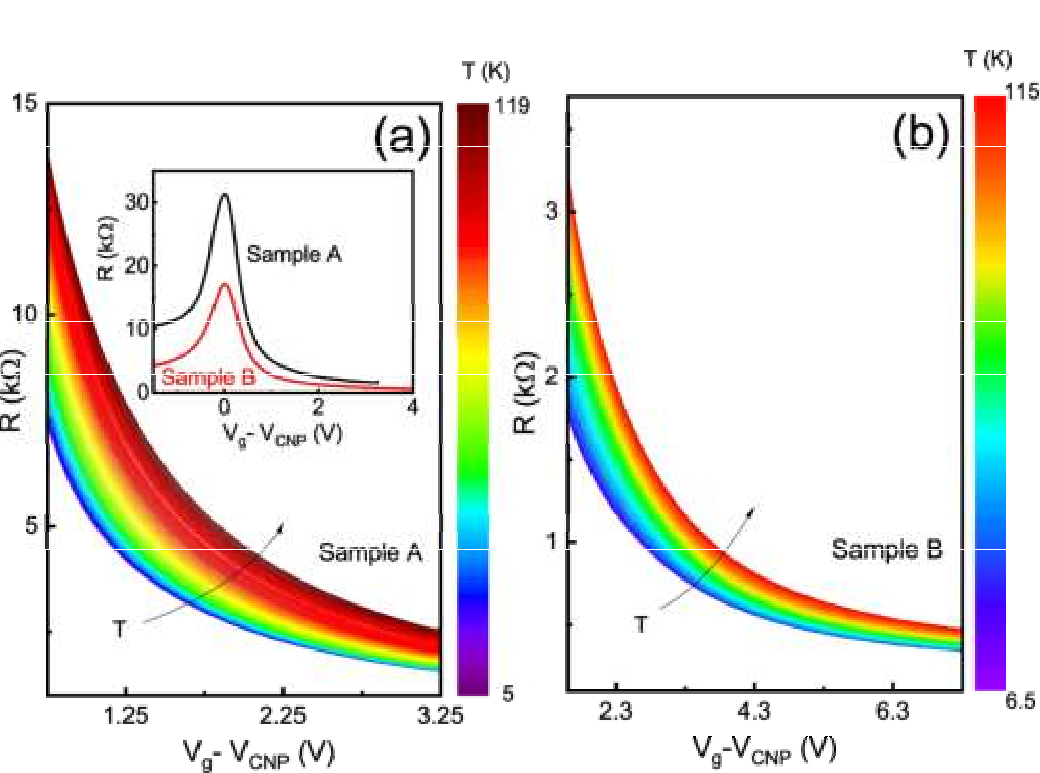}
\caption{(Color online) Resistance as a function of the gate voltage at different temperatures  for sample A (a)  and sample B (b) for electron side of the energy spectrum. Insert to Fig 1 : Resistance as a function of the gate voltage at
 for two gapless HgTe  quantum wells, T=5K.}
\end{figure}
\begin{figure}
\includegraphics[width=9cm]{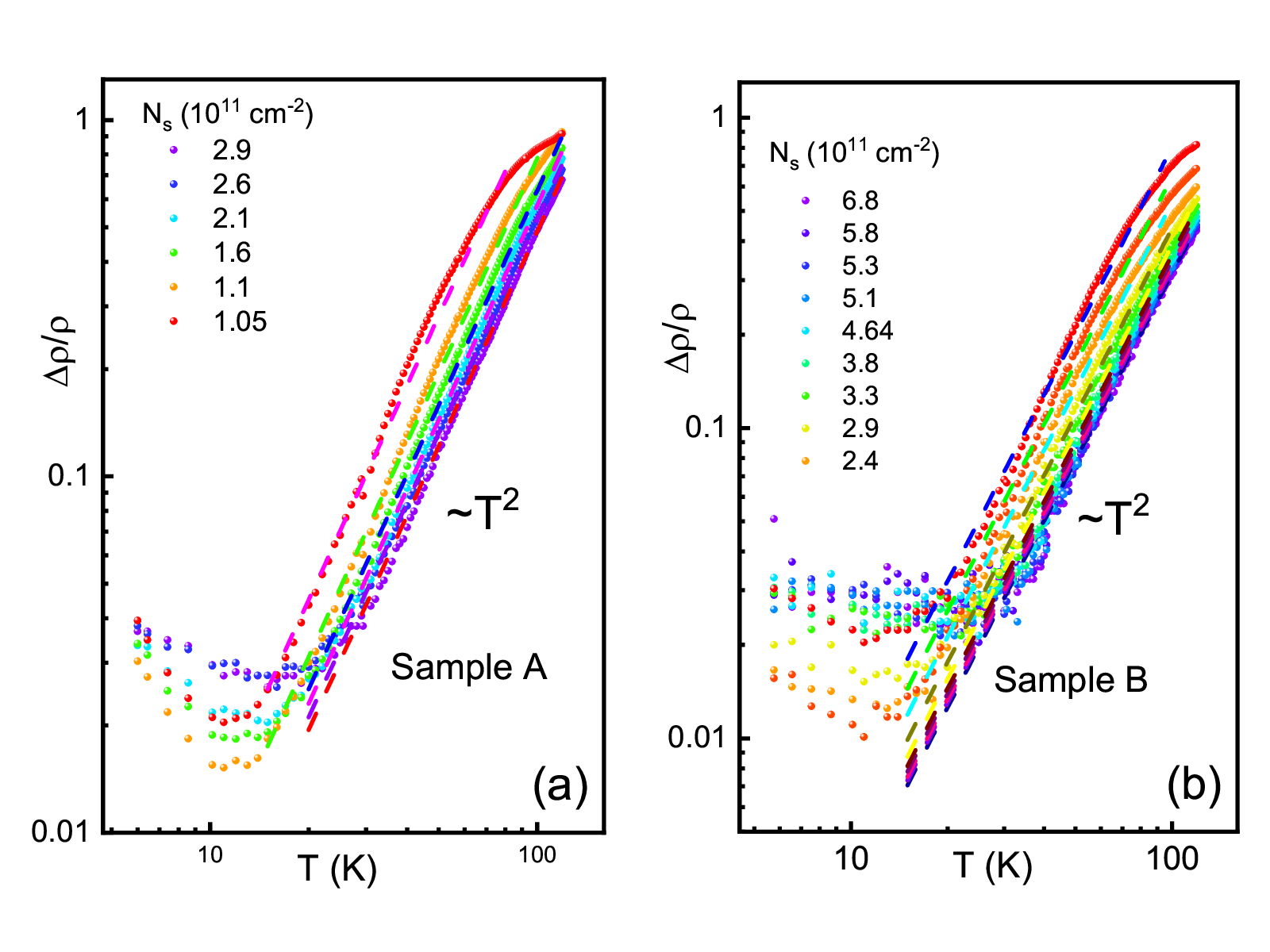}
\caption{(Color online) Excess resistivity $\Delta \rho(T)=(\rho(T)-\rho(T=5K))/\rho(T=5K)$ as a function of the temperature for various densities for sample A (a) and B (b). The dashes show $T^{2}$ dependence, calculated from Eq.\eqref{resistivity_correction}.}
\end{figure}


\textit{Experiment.---}HgTe-based quantum wells have attracted considerable attention due to their ability to generate a range of unconventional quantum materials, which are influenced by the quantum well's thickness. Notably, the energy spectrum of a gapless HgTe quantum well with a width of 6.3-6.5 nm features a single-valley Dirac cone at the center of the Brillouin zone. This characteristic makes the system quite similar to graphene. However, due to the single valley structure, the electronic properties differ significantly from those of the multi-valley graphene. In this HgTe well, Dirac fermions exhibit a linear energy spectrum for both electrons and holes, given by $\varepsilon_{p} = \pm v|p|$, where the Fermi velocity is $v = 7\times10^7 cm/s = c/430$ (with $c$ being the speed of light) and $k$ representing the momentum \cite{buttner}.

The inset of Fig. 1 illustrates the variation in resistivity as a function of gate voltage for samples A and B at a temperature of 5 K. The resistance displays a pronounced peak at the charge neutrality point, corresponding to the zero-energy level. This behavior is characteristic of gapless semiconductors, such as graphene. The maximum electron density observed corresponds to a Fermi energy of approximately 150 meV.

We fabricated  quantum wells using $HgTe/Cd_{x}Hg_{1-x}Te$ material with a $[013]$ surface orientation.  The wells had equal widths, with $d_0$ measuring 6.7 nm. The devices used in this study were multiterminal bars with three consecutive segments, each 50 $\mu m$ wide, and varying lengths of 100 $\mu m$, 250 $\mu m$, and 100 $\mu m$ \cite{suppl}. A 200 nm $SiO_{2}$ dielectric layer was deposited on the sample surface, which was then covered by a TiAu gate. The density variation with gate voltage was estimated to be approximately $0.9 \times 10^{11} cm^{-2}/V$, calculated from the dielectric thickness and Hall measurements.
Figures 1b and 1b display the variation of resistance with gate voltage across a broad temperature range, specifically for the gate voltage interval corresponding to the electronic part of the spectrum. The plot shows a notable increase in resistance as the temperature rises, with one exception: in the temperature range $5  \text{K} < T < 20  \text{K}$, a reduction in resistance with increasing temperature is observed. This phenomenon can be attributed to the weak localization effect, which has been previously reported in the study \cite{kozlov, suppl}.

To further explore the temperature-dependent behavior of resistance (or resistivity), we calculate the the ratio of the excess resistivity to resistivity at T=5 K, denoted as $\Delta \rho(T)/\rho(T=5 K) = (\rho(T) - \rho(T=5 \text{K}))/\rho(T= 5 K)$. Figure 2 presents the excess resistance for different electron densities across a wide temperature range for both samples A and B. The main feature of the experimental dependencies shown in Fig.2 is the presence of two distinct regimes with fundamentally different temperature behaviors, separated by an excess resistance minimum at $T^* \sim (10\text{--}20)\,\mathrm{K}$. The value of $T^*$ varies slightly between different samples and electron densities.

In the low-temperature regime ($T < T^*$), the excess resistance increases slightly as temperature approaches zero; we attribute this behavior to weak localization effects. Conversely, in the high-temperature regime ($T > T^*$), the excess resistivity follows a $T^{2}$ dependence across all measured electron densities. This behavior suggests a dominant contribution from electron-electron scattering to the excess resistance.  It's important to note that phonon scattering results in a linear temperature dependence \cite{melezhik}. An evaluation of the scattering time indicates that it is significantly smaller than what we observed in our temperature range. If this were not the case, we would expect to see strong deviations of $\rho (T)$ from the $T^2$ dependence, which is not observed (see Fig. 2). 

A comparison involves matching the experimental behavior with the theoretical expression in Eq.~\eqref{int}. Notably, Eq.~\eqref{int} also exhibits two distinct temperature regimes with different scaling behaviors. As we mentioned above we find that: 
At low temperatures ($T \ll T^*$), Eq.~\eqref{int} gives a correction $\sigma_{xx} \sim T^4 \ln(2\mu/T)$. As $T \to 0$, this contribution becomes negligible, leaving only the dominant weak localization behavior observed experimentally.
In contrast, at higher temperatures, $T\gg T^*$, Eq.\eqref{int} yields
\begin{gather}
\sigma_{xx}=-\frac{e^2}{6\hbar}
\left(\frac{e^2}{\varepsilon\hbar v}\right)^2
\left(\frac{T\tau_i}{\hbar}\right)^2.
\end{gather}

\begin{gather}\label{resistivity_correction}
\frac{\delta\rho}{\rho_0}=-\frac{\sigma_{xx}}{\sigma_0}=
\frac{2}{3}\left(\frac{\hbar}{\tau_{ee}}\right)\left(\frac{\tau_i}{\hbar}\right),
\end{gather}
where $\frac{\hbar}{\tau_{ee}}=\left(\frac{Ce^2}{\varepsilon\hbar v}\right)^2
\frac{T^2}{\mu}=(C \alpha_{ee})^2\frac{T^2}{\mu}$ is an inverse electron-electron scattering time and $\alpha_{ee}=3.2/\epsilon$, where $C$ is parameter considered below. 
 Expression Eq.\eqref{resistivity_correction} is the inter-particle collision correction to the system resistivity for the case of degenerate electrons with linear Dirac spectrum. Note, however, that the simplified form of the Coulomb potential does not account for either the finite thickness of the quantum well or dynamic screening effects within the random phase approximation (RPA). The relaxation of various perturbation types in a 2D Fermi gas was theoretically analyzed in Ref.~\cite{alekseev}. Calculations of both weak- and strong-interaction limits demonstrated that the scattering time $\tau_{ee}$ depends critically on the interaction parameter $r_s = 1/(\sqrt{\pi n} a_B)$, where $a_B$ is the Bohr radius. To achieve better agreement with theoretical predictions, we introduce an additional fitting parameter $C$ in our analysis.

We compare Eq.\eqref{resistivity_correction} with the experimental curves shown in Figure 2 by adjusting the parameters $\tau_{i}$ and the Coulomb interaction constant $C$ in the expression (6).
We take $\epsilon =10$. Figure 2 shows the theoretical curves for the parameters listed in Table I. It can be observed that the experimental data closely follows the expected trend $\Delta \rho(T)/\rho(T=5 K) \sim T^{2}$. For a system of massless Dirac fermions, the parameter $C$ is given by \cite{narozhny2}: $C \approx 4[ln(\mu /kT )]^{1/2}$, which approximately aligns with our extracted parameter for high-density conditions. It is important to highlight that, despite the electron-electron scattering time being closely related to hydrodynamic flow in an electron liquid, this approach predicts a temperature dependence of resistivity $\rho \sim T^{-2}$ \cite{gurzhi, spivak, suppl}, which is inconsistent with our experimental results.

\begin{table}[b!]
\centering
\begin{tabular}{|l|l|l|l||l||l|}
\hline
sample & $\tau_{i}$ $(10^{-12} s)$ & $\mu (meV)$ & C \\
\hline
\hline
A&   0.56 & 62 & 2.6 \\
\hline
A&   0.47 & 51 & 3 \\
\hline
A&   0.41 & 45 & 3.2 \\
\hline
B&    1.7 & 92.8 & 1.5 \\
\hline
B&    1.4 &70.8  & 1.6 \\
\hline
B&    1 & 56.8 & 2 \\
\hline

\hline
\end{tabular}
\caption{\label{tab1} Fitting parameters in Eq.\eqref{resistivity_correction}  for samples A and B. }
\end{table}

A theoretical estimate of the characteristic temperature $T^*$  using material parameters from the experimental samples, yields $T^* \sim (150\text{--}170)\,\mathrm{K}$. This value exceeds the experimentally observed range of $T^* \sim (10\text{--}20)\,\mathrm{K}$ by an order of magnitude. We attribute this discrepancy to limitations in our theoretical model, which assumes perfect two-dimensionality and neglects finite sample width effects. These approximations lead to an incomplete treatment of screening phenomena, ultimately resulting in an overestimated $T^*$ value. It is worth noting that additional effects, such as the renormalization of the Fermi velocity due to electron-electron (e-e) interactions, may also play a role, as theoretically predicted in various models (for a review, see \cite{kotov}). Crucially, however, this renormalization is expected to be most pronounced near the Dirac point. The experimentally observed logarithmic velocity renormalization \cite{elias} is in agreement with theoretical predictions, offering direct evidence that long-range e-e interactions can significantly alter the Dirac cone structure in the vicinity of the Dirac point. However, our study focuses on energy regimes far from the Dirac point, where we believe such renormalization effects are not significant and are unlikely to influence transport properties.
Although we did not observe a $T^4$ dependence in our HgTe quantum well, we expect graphene monolayers to better approximate the model interaction due to their screened Coulomb potential, with a characteristic temperature $T^* \approx 100 K$. Consequently, the $T^4$ contribution should be more readily observable at low temperatures in graphene. Despite the significant interest in Dirac materials, temperature-dependence studies have primarily focused on the Dirac point regime, where electron-hole plasma effects dominate \cite{xin}. A more systematic investigation of the temperature scaling across different carrier densities could help clarify the nature of the Coulomb potential in these systems.

In conclusion, we have theoretically and experimentally studied the T dependent corrections to the resistivity due to electron-electron interactions  in systems with systems with a $p$-linear spectrum. These effects are absent in Galilean-invariant systems with a parabolic spectrum. We believe that electronic transport phenomena are far from fully understood, and our research demonstrates that electron-electron dominated transport is significantly influenced by material properties, including the shape of the Fermi surface and the dispersion relation. Among the Dirac materials graphene, including its moiré and twisted forms, remains a fascinating and promising subject for investigating the contribution of electron-electron (e-e) scattering to resistivity \cite{efetov}. Transition metal dichalcogenide (TMD) systems have also gained significant attention recently for studying e-e interaction effects. However the linear spectrum approach in TMDs requires extremely high electron densities, which poses experimental challenges \cite{kis}. A particularly promising Dirac cone system has been observed in thin films of $Cd_3As_2$ \cite{stemmer}. Three-dimensional topological insulators host Dirac cone states on their surfaces, making them a fascinating platform for investigating the interplay between electron-electron interactions and topological effects. Understanding this relationship could provide deeper insights into their transport properties \cite{hasan}.

The financial support of this work by Sao Paulo  Research Foundation (FAPESP) Grant No. 2019/16736-2 and No. 2021/12470-8, the National Council for Scientific and Technological Development (CNPq)  Ministry of Science and Higher Education of the Russian Federation, and Foundation for the Advancement of Theoretical Physics and Mathematics "BASIS" is acknowledge. HgTe quantum wells growth and preliminary transport measurements are supported by Russian Science Foundation (Grant No. 23-72-30003 ).

\clearpage
\appendix
\section*{SUPPLEMENTAl MATERIAL: Resistivity of non-Galilean invariant two dimensional Dirac system}

\section*{Supplemental Material Abstract}
In this supplemental material, we present the detailed solution of the Boltzmann equation for electron-impurity and electron-electron scattering, which is part of the theoretical section of the paper. For the experimental section, we provide detailed information on the sample characteristics and the measurement methods used.

\date{\today}

\maketitle

\vspace{20pt}




\section*{Solution of Boltzman equation for electron-impurity and electron-electron scattering}
The Boltzmann equation, describing (i) the electron scattering on impurities and (ii) the electron-electron scattering, reads ($e>0$): 
\begin{gather}\label{q1}
-{\bf F}\cdot\frac{\partial f_{\bf p}}{\partial {\bf p}}
=
-\frac{f_{{\bf p}}-n_{{\bf p}}}{\tau_i}
+Q_{ee}\{f_{\bf p}\},
\end{gather}
where the first term on r.h.s. is an electron-impurity collision integral taken in $\tau_i$-approximation, the second term is an electron-electron collision integral. $n_{{\bf p}}=\left[\exp\left(\frac{\epsilon_\mathbf{p}-\mu}{T}\right)+1\right]^{-1}$ is the equilibrium electron distribution function, $\epsilon_p=vp$ is electron dispersion. 
We expand the nonequilibrium distribution functions into the first order with respect to the external force ${\bf F}$ as $f_{\textbf{p}(\textbf{k})}=n_{\textbf{p}}+\delta f_{\textbf{p}}$, where $\delta f_{\textbf{p}}=\delta f^{(0)}_{\textbf{p}}
+\delta f^{(1)}_{\textbf{p}}$ are the zero- and first order iteration corrections with respect to small inter-particle interaction term $Q_{ee}$. Deploying the method of successive iterations results in:
\begin{eqnarray}
\delta f^{(0)}_{\bf p}=
\tau_i({\bf F}\cdot{\bf v}_{\bf p})n'_{\bf p}=\tau_i\phi_{\bf p}n'_{\bf p}
\end{eqnarray}
\begin{eqnarray}
\delta f^{(1)}_{\bf p}
=
\tau_iQ_{ee}\{\delta f^{(0)}_{\bf p}\},
\end{eqnarray}
where the particle-particle collision integral being linearized reads
\begin{widetext}
\begin{eqnarray}
\label{EqQee1}
Q_{ee}\left\{\delta f_{\bf p}\right\}&=&-2\pi\sum_{\mathbf{p}',\mathbf{k},\mathbf{k}'}
|U_{{\bf p}'-{\bf p}}|^2
\delta(\epsilon_{{\bf k}'}+\epsilon_{{\bf p}'}-\epsilon_{{\bf k}}-\epsilon_{{\bf p}})
\delta_{{\bf k}'+{\bf p}'-{\bf k}-{\bf p}}
\\
\nonumber
&&~~~~~\times\Bigl[\delta f_{\bf p}[(1-n_{\bf k})n_{{\bf k}'}n_{{\bf p}'}+n_{\bf k}(1-n_{{\bf k}'})(1-n_{{\bf p}'})]
-\delta f_{{\bf p}'}[(1-n_{{\bf k}})(1-n_{{\bf p}})n_{{\bf k}'}+n_{\bf k}n_{\bf p}(1-n_{{\bf k}'})]\\
\nonumber
&&~~~~~~~+\delta f_{\bf k}[(1-n_{\bf p})n_{{\bf k}'}n_{{\bf p}'}+n_{\bf p}(1-n_{{\bf k}'})(1-n_{{\bf p}'})]
-\delta f_{{\bf k}'}[(1-n_{{\bf k}})(1-n_{{\bf p}})n_{{\bf p}'}+n_{\bf k}n_{\bf p}(1-n_{{\bf p}'})]\Bigr].
\end{eqnarray}
\end{widetext}
Electric current density correction reads as 
\begin{eqnarray}
{\bf j}=-
e
\sum_{\bf p}{\bf v}_{\bf p}\delta f^{(1)}_{\bf p}=-e\tau_i\sum_{\bf p}{\bf v}_{\bf p}
Q_{ee}\{\delta f^{(0)}_{\bf p}\bigr).
\end{eqnarray}

Substituting the ansatz $\delta f_\mathbf{p}^{(1)}=\tau_i\phi_\mathbf{p}n_\mathbf{p}'$ in Eq.~\eqref{EqQee1}, one finds:
\begin{widetext}
\begin{gather}
j_\alpha=-2\pi e\tau_i^2\sum_{\bf p}
v_\alpha({\bf p})
\sum_{\mathbf{p}',\mathbf{k}',\mathbf{k}}|U_{\mathbf{p}-\mathbf{p}'}|^2(\phi_\mathbf{p}-\phi_{\mathbf{p}'}+\phi_\mathbf{k}-\phi_{\mathbf{k}'})
(n_\mathbf{p}-n_{\mathbf{p}'})
(n_\mathbf{k}-n_{\mathbf{k}'})\\
\nonumber\times
\delta(\mathbf{p}+\mathbf{k}-\mathbf{p}'-\mathbf{k}')
\int d\omega \frac{dN_\omega}{d\omega}
\delta(\epsilon_{\mathbf{k}'}-\epsilon_\mathbf{k}-\omega)
\delta(\epsilon_{\mathbf{p}'}-\epsilon_\mathbf{p}+\omega),
\end{gather}

where

\begin{eqnarray}
\label{EqNw}
N_\omega=\frac{1}{e^{\omega/T}-1},\,\,\,\,\textrm{and thus},\,\
\frac{\partial N}{\partial \omega}=\frac{N_{-\omega}N_{\omega}}{T}=-\frac{N_{\omega}(1+N_\omega)}{T}=-\frac{1}{4T\sinh^2\left(\frac{\omega}{2T}\right)},
\end{eqnarray}
\end{widetext}
and $\phi_{\bf p}=({\bf F}\cdot{\bf v}_{\bf p})$. Taking onto account the symmetry of current density expression, one finds the final expression for the current correction
\begin{widetext}
\begin{gather}\nonumber
j_\alpha=2\pi e\tau_i^2\sum_{{\bf p},{\bf k},{\bf p}',{\bf k}'}
({\bf F}\cdot{\bf v}_{\bf p})
[{\bf v}_{\bf p}
+{\bf v}_{\bf k}
-{\bf v}_{{\bf p}'}
-{\bf v}_{{\bf k}'}]
|U_{\mathbf{p}-\mathbf{p}'}|^2
(n_\mathbf{p}-n_{\mathbf{p}'})
(n_\mathbf{k}-n_{\mathbf{k}'})\\
\times
\delta(\mathbf{p}+\mathbf{k}-\mathbf{p}'-\mathbf{k}')
\int  \frac{d\omega}{4T\sinh^2\left(\frac{\omega}{2T}\right)}
\delta(\epsilon_{\mathbf{k}'}-\epsilon_\mathbf{k}-\omega)
\delta(\epsilon_{\mathbf{p}'}-\epsilon_\mathbf{p}+\omega).
\label{current_general}
\end{gather}
\end{widetext}
Introducing the transferred momentum under interparticle collisions,
${\bf q} ={\bf k}'-{\bf k}={\bf p}-{\bf p}'$, one finds
\begin{widetext}
\begin{gather}\nonumber
{\bf j}=2\pi e\tau_i^2\sum_{{\bf p},{\bf k},{\bf q}}
({\bf F}\cdot{\bf v}_{\bf p})
[{\bf v}_{\bf p}
+{\bf v}_{\bf k}
-{\bf v}_{{\bf p}-{\bf q}}
-{\bf v}_{{\bf k}+{\bf q}}]
|U_{\mathbf{q}}|^2
(n_\mathbf{p}-n_{\mathbf{p}-\mathbf{q}})
(n_\mathbf{k}-n_{\mathbf{k}+\mathbf{q}})\\
\times
\int  \frac{d\omega}{4T\sinh^2\left(\frac{\omega}{2T}\right)}
\delta(\epsilon_{\mathbf{k}+\mathbf{q}}-\epsilon_\mathbf{k}-\omega)
\delta(\epsilon_{\mathbf{p}-\mathbf{q}}-\epsilon_\mathbf{p}+\omega).
\label{current_general}
\end{gather}
\end{widetext}
This is a general expression for the current density correction caused by the interparticle collisions. Based on this expression, one considers below the case of degenerate electron gas.


\subsection{Degenerate electron gas case}
Expanding the distribution functions in Eq.\eqref{current_general} as
\begin{eqnarray}
n_\mathbf{p}-n_{\mathbf{p}-\mathbf{q}}=\omega n_\mathbf{p}'
,\,\,\,
n_\mathbf{k}-n_{\mathbf{k}+\mathbf{q}}=-\omega n_\mathbf{k}'
\end{eqnarray}
and using the relations
\begin{widetext}
\begin{gather}
{\bf v}_{\bf p}
+{\bf v}_{\bf k}
-{\bf v}_{{\bf p}-{\bf q}}
-{\bf v}_{{\bf k}+{\bf q}}]=
v^2\left[
\frac{p_\alpha}{\epsilon_{\bf p}}
-\frac{p_\alpha-q_\alpha}{\epsilon_{\bf p}-\omega}
+\frac{k_\alpha}{\epsilon_{\bf k}}
-\frac{k_\alpha+q_\alpha}{\epsilon_{\bf k}+\omega}
\right]=\\\nonumber
v^2\left[
\frac{q_\alpha\epsilon_{\bf p}-p_\alpha\omega}{\epsilon_{\bf p}(\epsilon_{\bf p}-\omega)}
-\frac{q_\alpha\epsilon_{\bf k}-k_\alpha\omega}{\epsilon_{\bf k}(\epsilon_{\bf k}+\omega)}
\right],
\end{gather}
\end{widetext}
and
\begin{widetext}
\begin{gather}
\delta(\epsilon_{\mathbf{k}+\mathbf{q}}-\epsilon_\mathbf{k}-\omega)=
(\epsilon_{\mathbf{k}+\mathbf{q}}+\epsilon_\mathbf{k}+\omega)
\delta(\epsilon^2_{\mathbf{k}-\mathbf{q}}-(\epsilon_\mathbf{k}+\omega)^2)=
2(\epsilon_\mathbf{k}+\omega)
\delta(\epsilon^2_{\mathbf{k}-\mathbf{q}}-(\epsilon_\mathbf{k}+\omega)^2),\\
\delta(\epsilon_{\mathbf{p}-\mathbf{q}}-\epsilon_\mathbf{p}+\omega)=
(\epsilon_{\mathbf{p}-\mathbf{q}}+\epsilon_\mathbf{p}-\omega)
\delta(\epsilon^2_{\mathbf{p}-\mathbf{q}}-(\epsilon_\mathbf{p}-\omega)^2)=
2(\epsilon_\mathbf{p}-\omega)
\delta(\epsilon^2_{\mathbf{p}-\mathbf{q}}-(\epsilon_\mathbf{p}-\omega)^2),
\end{gather}

the current can be written as

\begin{gather}
j_\alpha=-
2\pi e\tau_i^2\frac{(2v)^2}{4T}
\sum_{{\bf p},{\bf k},{\bf q}}
|U_{\bf q}|^2 n_{\bf p}'n_{\bf k}'
\int\frac{\omega^2d\omega}{\sinh^2(\omega/2T)}
({\bf F}\cdot{\bf v}_{\bf p})
\left[
\frac{(q_\alpha\epsilon_{\bf p}-p_\alpha\omega)(\epsilon_{\bf k}+\omega)}
{\epsilon_{\bf p}}
-\frac{(q_\alpha\epsilon_{\bf k}-k_\alpha\omega)(\epsilon_{\bf p}-\omega)}{\epsilon_{\bf k}}
\right]\times\\\nonumber\times
\delta(\epsilon^2_{\mathbf{k}+\mathbf{q}}-(\epsilon_\mathbf{k}+\omega)^2)
\delta(\epsilon^2_{\mathbf{p}-\mathbf{q}}-(\epsilon_\mathbf{p}-\omega)^2)
\end{gather}
\end{widetext}
The relations $n_{\bf p}'n_{\bf k}'=\delta(\epsilon_{\bf p}-\mu)\delta(\epsilon_{\bf k}-\mu)$ that hold for degenerate electron gas, allow the integration over absolute values of momenta $\mu=vp_0=vk=vp$. Thus, we get for conductivity
\begin{widetext}
\begin{gather}
\sigma_{xx}=-\frac{(e\tau_iv\mu)^2v}{2\pi T v^4}
\int\limits_0^\infty\frac{qdq}{2\pi}
\int\limits_0^{2\pi}\frac{d\alpha}{2\pi}
|U_{\bf q}|^2 
\int\limits_0^{2\pi}\frac{d\phi_p}{2\pi}
\int\limits_0^{2\pi}\frac{d\phi_k}{2\pi}
\int\frac{\omega^2d\omega}{\sinh^2(\omega/2T)}
\cos(\phi_p+\alpha)\\\nonumber
\left[
\frac{[q\mu\cos(\alpha)-\omega p_0\cos(\phi_p+\alpha)](\mu+\omega)}
{\mu}
-\frac{[q\mu\cos(\alpha)-\omega p_0\cos(\phi_k+\alpha)](\mu-\omega)}{\mu}
\right]\times\\\nonumber\times
\delta[2\mu\epsilon_q\cos\phi_k+\mu^2+\epsilon_q^2-(\mu+\omega)^2]
\delta[2\mu\epsilon_q\cos\phi_q-\mu^2-\epsilon_q^2+(\mu-\omega)^2].
\end{gather}
\end{widetext}
Here, $\epsilon_q=vq$. Integrating over $\alpha$, we get
\begin{widetext}
\begin{gather}
\sigma_{xx}=-\frac{(e\tau_iv\mu)^2v}{2\pi T v^4}
\int\limits_0^\infty\frac{qdq}{2\pi}
\frac{1}{2}
|U_{\bf q}|^2 
\int\limits_0^{2\pi}\frac{d\phi_p}{2\pi}
\int\limits_0^{2\pi}\frac{d\phi_k}{2\pi}
\int\frac{\omega^2d\omega}{\sinh^2(\omega/2T)}
\\\nonumber
\left[
\frac{[q\mu\cos(\phi_p)-\omega p_0](\mu+\omega)}
{\mu}
-\frac{[q\mu\cos(\phi_p)-\omega p_0\cos(\phi_k-\phi_p)](\mu-\omega)}{\mu}
\right]\times\\\nonumber\times
\delta[2\mu\epsilon_q\cos\phi_k+\mu^2+\epsilon_q^2-(\mu+\omega)^2]
\delta[2\mu\epsilon_p\cos\phi_p-\mu^2-\epsilon_q^2+(\mu-\omega)^2].
\end{gather}
\end{widetext}
Now expressing the angles from delta-functions as
\begin{gather}
\cos\phi_k=\frac{(\mu+\omega)^2-\mu^2-\epsilon^2_q}{2\mu\epsilon_q},\\\nonumber
\cos\phi_p=\frac{\mu^2+\epsilon^2_q-(\mu-\omega)^2}{2\mu\epsilon_q},
\end{gather}
and using the integral
\begin{gather}
\int\limits_0^{2\pi}\frac{d\phi}{2\pi}\delta(a\cos\phi-b)
=\frac{1}{\pi}\frac{\theta[a^2-b^2]}{\sqrt{a^2-b^2}},
\end{gather}
One finds
\begin{widetext}
\begin{gather}
\sigma_{xx}=-\frac{(e\tau_iv\mu)^2v}{2\pi T 2\pi^2v^4}
\int\limits_0^\infty\frac{qdq}{2\pi}
|U_{\bf q}|^2 
\int\frac{\omega^2d\omega}{\sinh^2(\omega/2T)}\times
\\\nonumber
\times\frac{1}{\mu}\left[
\left(q\mu\frac{\mu^2+\epsilon^2_q-(\mu-\omega)^2}{2\mu\epsilon_q}-\omega p_0\right)(\mu+\omega)
-\left(q\mu\frac{\mu^2+\epsilon^2_q-(\mu-\omega)^2}{2\mu\epsilon_q}-\omega p_0\frac{\mu^2+\epsilon^2_q-(\mu-\omega)^2}{2\mu\epsilon_q}\frac{(\mu+\omega)^2-\mu^2-\epsilon^2_q}{2\mu\epsilon_q}\right)(\mu-\omega)
\right]\times\\\nonumber\times
\frac{1}{\sqrt{(\epsilon_q^2-\omega^2)[(2\mu+\omega)^2-\epsilon_q^2]}}
\frac{1}{\sqrt{(\epsilon_q^2-\omega^2)[(2\mu-\omega)^2-\epsilon_q^2]}}.
\end{gather}
\end{widetext}

\begin{figure*}

\includegraphics[width=12cm]{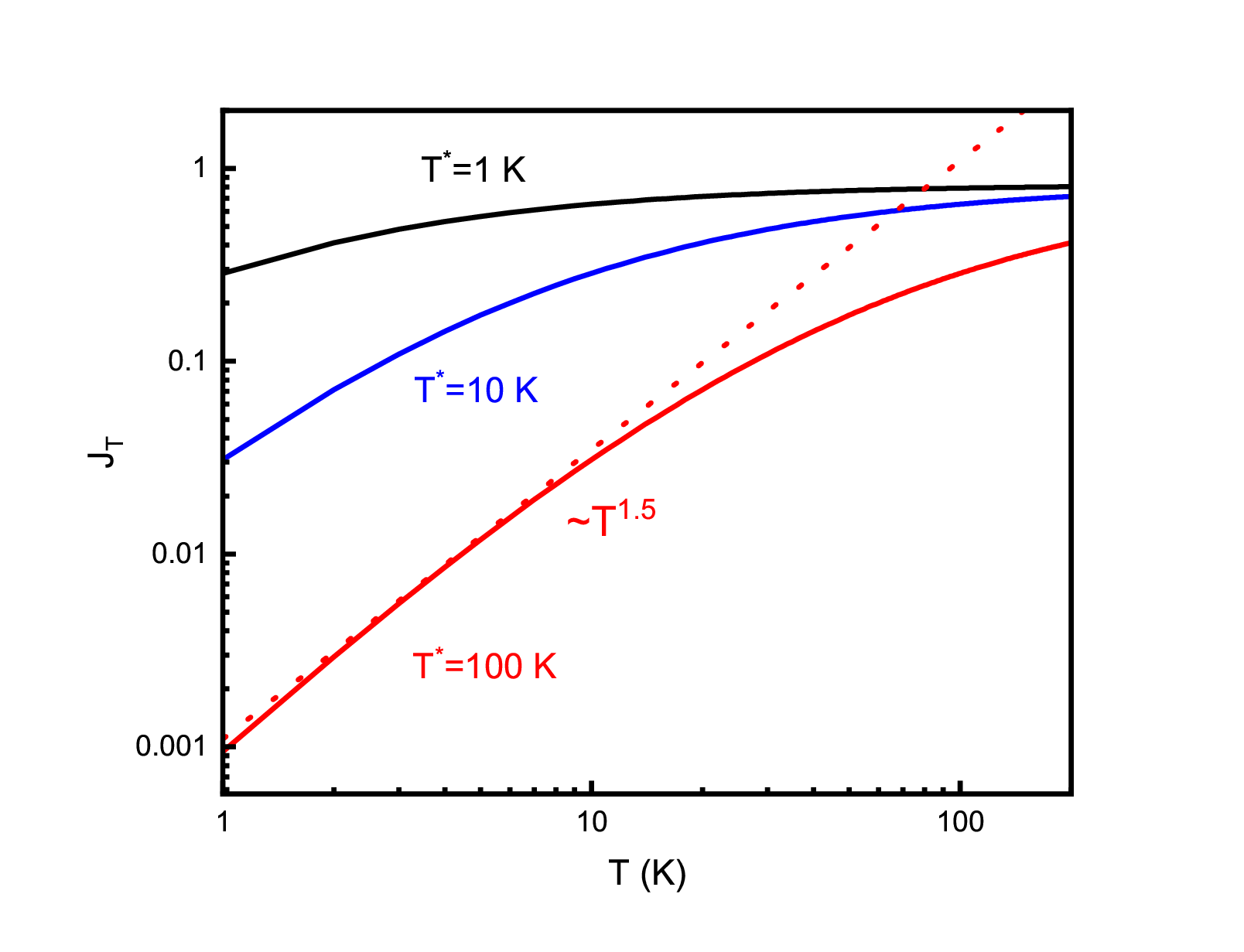}
\caption{The temperature dependence of factor $\mathfrak{J}_T$  (Eq.~(24)) for three different values of $T^{\star}$: 1, 10, and 100\,K. Dots represent the $T^{1.5}$ dependence shown for comparison.}
\label{integral}
\end{figure*}
It is possible to simplify the expression
\begin{widetext}
\begin{gather}
\left[
\left(q\mu\frac{\mu^2+\epsilon^2_q-(\mu-\omega)^2}{2\mu\epsilon_q}-\omega p_0\right)(\mu+\omega)
-\left(q\mu\frac{\mu^2+\epsilon^2_q-(\mu-\omega)^2}{2\mu\epsilon_q}-\omega p_0\frac{\mu^2+\epsilon^2_q-(\mu-\omega)^2}{2\mu\epsilon_q}\frac{(\mu+\omega)^2-\mu^2-\epsilon^2_q}{2\mu\epsilon_q}\right)(\mu-\omega)
\right]=\\\nonumber
=\frac{p_0\omega(\epsilon_q^2-\omega^2)(-4\mu^3+4\mu^2\omega+\mu\omega^2-\omega^3+\epsilon_q^2(3\mu+\omega))}{4\mu^2\epsilon_q^2},
\end{gather}
\end{widetext}
and find the following expression for conductivity correction (keeping the largest $4\mu^2\omega$ term in numerator)
\begin{widetext}
\begin{gather}
\sigma_{xx}=-\frac{(e\tau_iv\mu)^2v}{2\pi T 2\pi^2v^4}
\int\limits_0^\infty\frac{qdq}{2\pi}
|U_{\bf q}|^2 
\int\limits_{-\infty}^{+\infty}\frac{\omega^2d\omega}{\sinh^2(\omega/2T)}
\left(\frac{p_0\omega^2}{\mu\epsilon_q^2}\right)
\frac{\theta[\epsilon_q^2-\omega^2]}{\sqrt{(2\mu+\omega)^2-\epsilon_q^2}}
\frac{1}{\sqrt{(2\mu-\omega)^2-\epsilon_q^2}}.
\end{gather}
\end{widetext}
Keeping here only the even-in-$\omega$ terms, and leaving only $2\mu\gg(\omega,\epsilon_q)$ in square roots, one finds
\begin{widetext}
\begin{gather}
\sigma_{xx}=-\frac{(e\tau_iv\mu)^2v}{2\pi T 2\pi^2v^4 (2\mu)^2v}
\int\limits_0^\infty\frac{qdq}{2\pi}
|U_{\bf q}|^2 
\int\limits_{-\infty}^{+\infty}\frac{\omega^2d\omega}{\sinh^2(\omega/2T)}
\frac{\omega^2}{\epsilon^2_q}\theta[\epsilon_q^2-\omega^2]=\\\nonumber
-\frac{(e\tau_iv\mu)^2v}{2\pi T 2\pi^2v^4 (2\mu)^2v}
\int\limits_{-\infty}^{+\infty}\frac{\omega^4d\omega}{\sinh^2(\omega/2T)}
\int\limits_{|\omega|/v}^\infty\frac{qdq}{2\pi}
\frac{|U_{\bf q}|^2}{\epsilon^2_q}.
\end{gather}
\end{widetext}
Remind that $\epsilon_q=vq$. Considering the bare Coulomb potential $U_{\bf q}=2\pi e^2/\varepsilon q$, the correction to conductivity reads:
\begin{gather}\label{int}
\frac{\delta \rho}{\rho_0} = \frac{8}{\pi^2}\left(\frac{e^2}{\epsilon \hbar v} \right)^2 \frac{T}{\mu} \left( \frac{T\tau}{\hbar} \right) \mathfrak{J}_T
\end{gather}
\begin{gather}
\mathfrak{J}_T =\left( \frac{T}{T^*} \right)^2 \int_{0}^{\infty} \frac{x^4 \, dx}{\sinh^2(x)} \left[ \ln \left( 1 + \frac{T^*}{xT} \right) - \frac{1}{1 + xT / T^*} \right]
\end{gather}
where $T^{\star} = \hbar \nu q_s = \frac{\mu e^2}{\varepsilon \hbar \nu}$ is the characteristic temperature separating different temperature regimes. Figure~\ref{integral} shows the temperature dependence of Eq.~(24) for three different values of $T^{\star}$: 1, 10, and 100\,K. One can see that for $T^{\star} = 1\,$K, the factor $\mathfrak{J}_T$ is almost independent of temperature. In this case,
\begin{gather}
\sigma_{xx}=-\frac{e^2}{6\hbar}
\left(\frac{e^2}{\varepsilon\hbar v}\right)^2
\left(\frac{T\tau_i}{\hbar}\right)^2.
\end{gather}
Taking into account the bare Drude conductivity for particles with Dirac linear spectrum
$$\sigma_0=\frac{e^2}{4\hbar}\left(\frac{\mu\tau_i}{\hbar}\right),$$ 
the correction to resistivity can be expressed via the particle-particle correction to the conductivity as
\begin{gather}
\frac{\delta\rho}{\rho_0}=-\frac{\sigma_{xx}}{\sigma_0}=
\frac{2}{3}
\left(\frac{e^2}{\varepsilon\hbar v}\right)^2
\frac{T}{\mu}
\left(\frac{T\tau_i}{\hbar}\right)\sim\frac{\tau_{i}}{\tau_{ee}},
\end{gather}
if one introduces the e-e scattering time as
\begin{gather}
\frac{\hbar}{\tau_{ee}}=\left(\frac{e^2}{\varepsilon\hbar v}\right)^2
\frac{T^2}{\mu}.
\end{gather}

For a high value of the characteristic temperature $T^{\star} = 100\,\mathrm{K}$, as shown in Figure~\ref{integral}, the integral in Eq.~(24) becomes weakly temperature dependent below $T \approx 20\,\mathrm{K}$, with the factor $\mathfrak{J}_T \sim \left(\frac{T}{T^{\star}}\right)^2$. This leads to corrections of the form $\sigma_{xx} \sim T^4 \ln(2\mu/T)$. Note, however, that above 20 K T the dependence of the factor $\mathfrak{J}_T$ is almost saturating. 
For the characteristic temperature $T^{\star} = 10\,\mathrm{K}$, the factor $\mathfrak{J}_T$ exhibits temperature dependence only below $20\,\mathrm{K}$, while rapidly saturating to a constant value at higher temperatures.

\section*{Experimental details and methods}
\subsection{Sample description}
\begin{figure*}
\includegraphics[width=16cm]{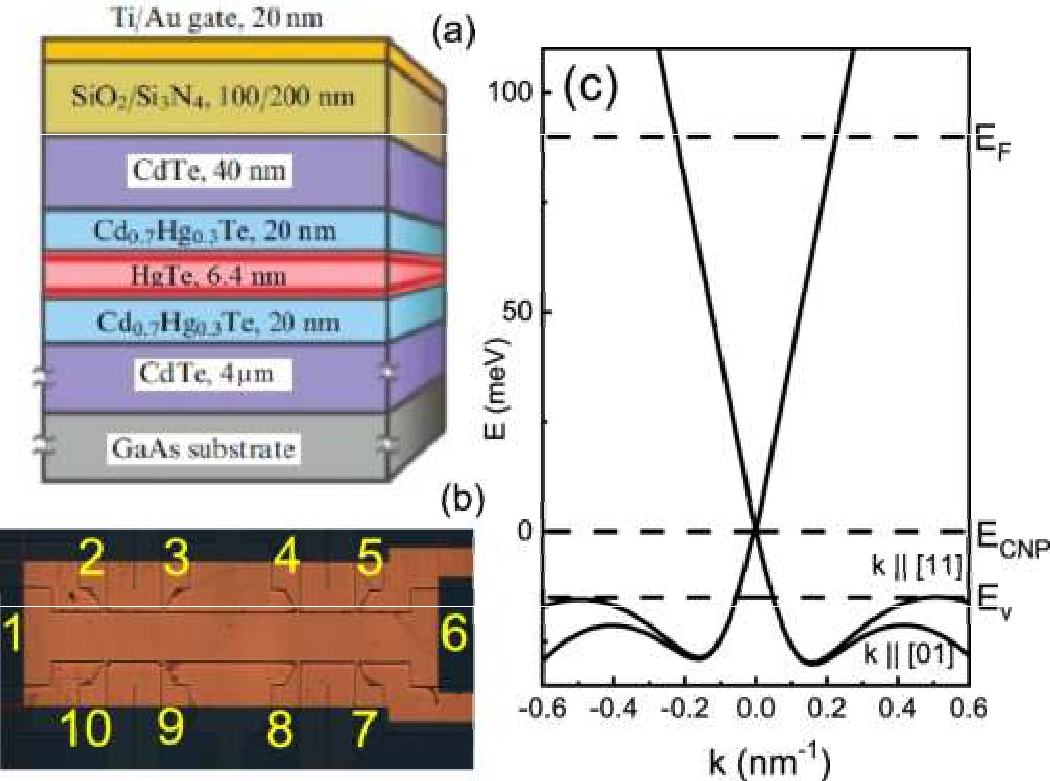}
\caption{ (a) Schematic of the transistor. (b) A top view of the sample. (c) Schematic representation of the energy spectrum of a 6.4-nm mercury telluride quantum well.}
\label{sample}
\end{figure*}

We fabricated  quantum wells using HgTe/Cd$x$Hg${1-x}$Te material with a $[013]$ surface orientation.  The wells had equal widths, with $d_0$ measuring 6.7 nm. The layer thickness was monitored during MBE growth via ellipsometry, achieving an accuracy within $\pm 0.3 nm$ (Fig.~\ref{sample}a).

The devices used in this study were multiterminal bars with three consecutive segments, each 50 $\mu m$ wide, and varying lengths of 100 $\mu m$, 250 $\mu m$, and 100 $\mu m$ (Fig.~\ref{sample}b). These devices featured nine contacts, which were created by indium bonding to the contact pads, precisely defined using lithography. Due to the relatively low growth temperature (around $180^{\circ} C$), the temperature during the contact fabrication process also remained low. Indium diffused vertically into each contact pad, forming an ohmic connection across all three quantum wells, with contact resistance ranging between 10 and 50 k$\Omega$ .

Throughout the AC measurements, we ensured that the reactive component of impedance remained below $5\%$ of the total impedance, confirming the effectiveness of the ohmic contacts. Additionally, the current-voltage (I-V) characteristics showed ohmic behavior at low voltages. A 200 nm $SiO_{2}$ dielectric layer was deposited on the sample surface, which was then covered by a TiAu gate. The density variation with gate voltage was estimated to be approximately $0.9 \times 10^{11} cm^{-2}/V$, calculated from the dielectric thickness and Hall measurements, as reported in previous studies using similar devices. Two samples from different substrates, labeled A and B, were analyzed.
\begin{table}[ht]
\centering
\begin{tabular}{|l|l|l|l||l||l||l|}
\hline
sample & $d$ (nm) & $V_{CNP}$ (V)& $\rho_{max}(h/e^{2})$& $\mu_{e} (V/cm^{2}s )$ \\
\hline
A&    6.3 & -1.25& 0.25 & 56.600\\
\hline
B&   6.4 & -4.3 & 0.12 &110.000\\
\hline
\end{tabular}
\caption{\label{tab1} Some of the typical parameters of the electron system in HgTe triple quantum well at T=4.2K.}
\end{table}
Table 1 provides the key parameters of the gapless HgTe quantum well used in this study. These parameters include the well width (d), the gate voltage associated with the Dirac point position ($V_{CNP}$), the resistivity ($\rho$) at the charge neutrality point (CNP), and the electron mobility ($\mu_e$), calculated as $1/\rho N_s$, where the total electron density ($N_s$) is set at $1 \times 10^{11}$ cm$^{-2}$.

\subsection{Measurements in magnetic field}

In this study, we examine both theoretically and experimentally the Dirac system in the deep \textit{n}-type regime, where only degenerate electrons are present. The position of the Fermi energy is indicated in Fig.~\ref{sample}c, demonstrating that the Fermi energy remains significantly larger than the thermal energy \(kT\) across the entire temperature range investigated in this experiment. Consequently, the electrons follow Fermi-Dirac statistics throughout the study.

In the main text, we measured the resistance as a function of carrier density far from the Dirac point in the absence of a magnetic field. It is important to emphasize that the observed temperature dependence of the resistance originates from electron-electron interactions in the single-subband regime. As discussed in the introduction, when two or more subbands are occupied, electron-electron scattering can significantly influence conductivity, leading to deviations from the observed behavior.

\begin{figure}
\includegraphics[width=9cm]{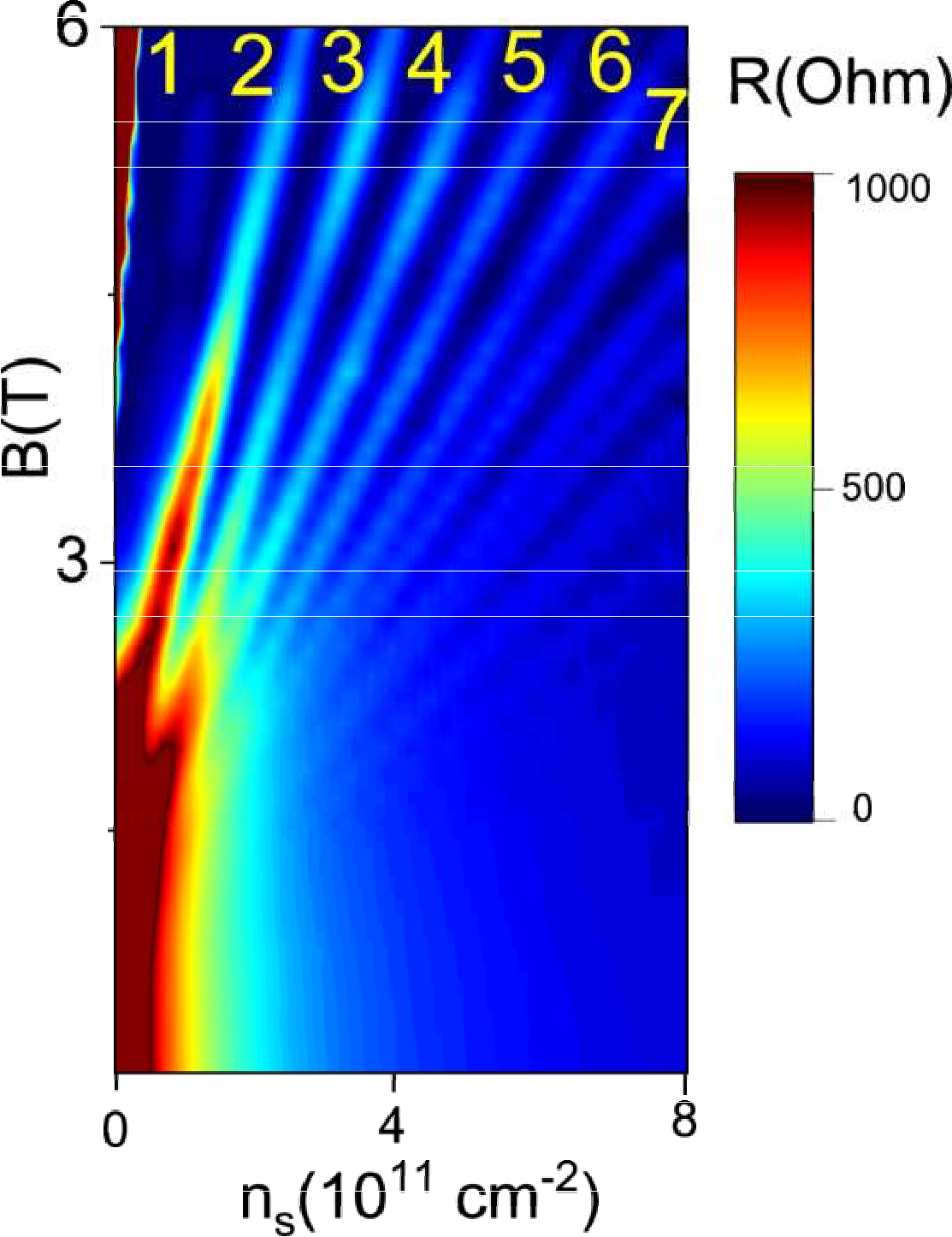}
\caption{ $R_xx(B, n_s)$ diagram, T=4.2 K. Filling factors $\nu$ determined from Hall resistance are labeled}
\label{Landau}
\end{figure}

\begin{figure}
\includegraphics[width=9cm]{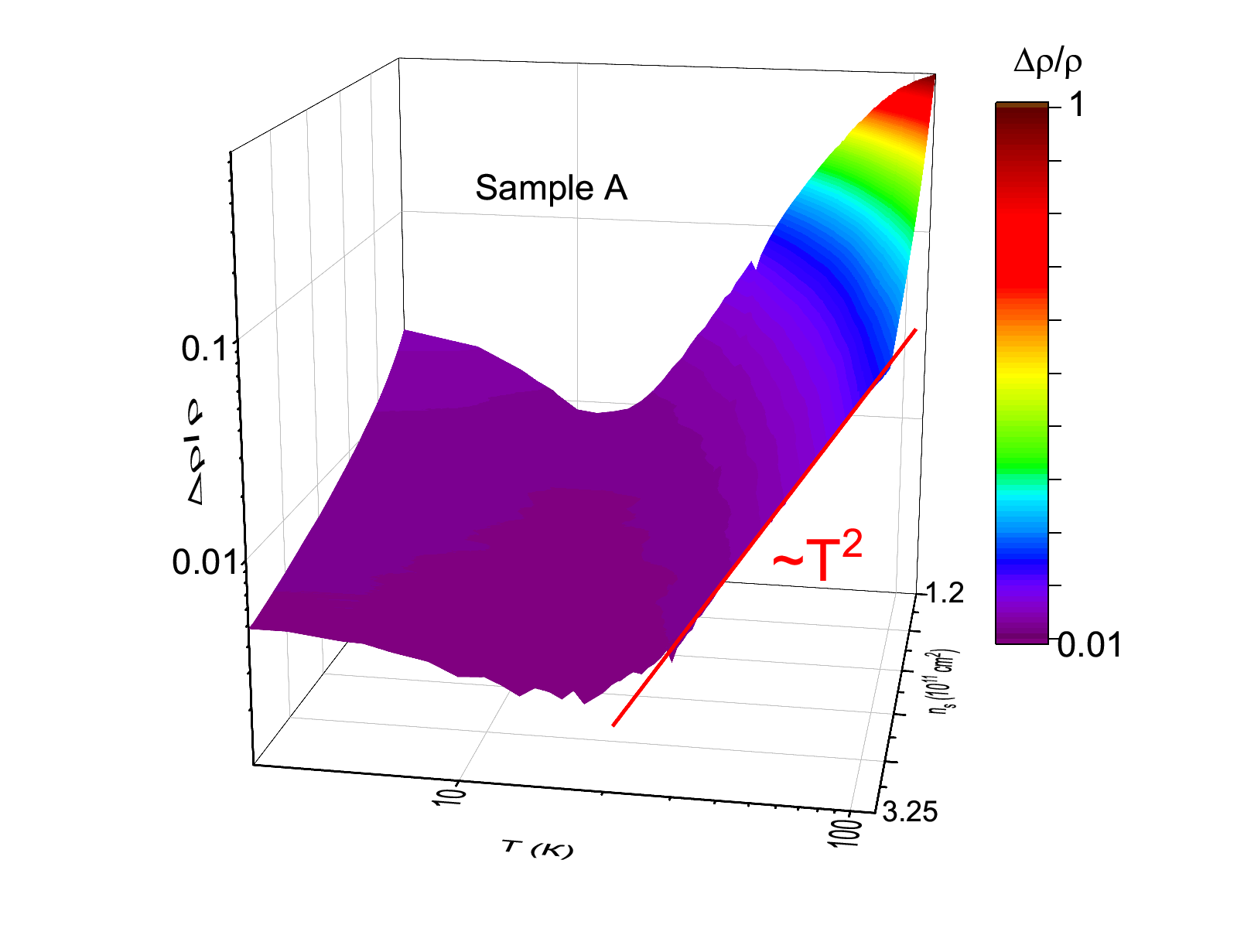}
\caption{ Excess of the resistivity as a function of the density and the temperature for sample A. Red line -$T^2$ dependence.}
\label{res1}
\end{figure}
\begin{figure}
\includegraphics[width=9cm]{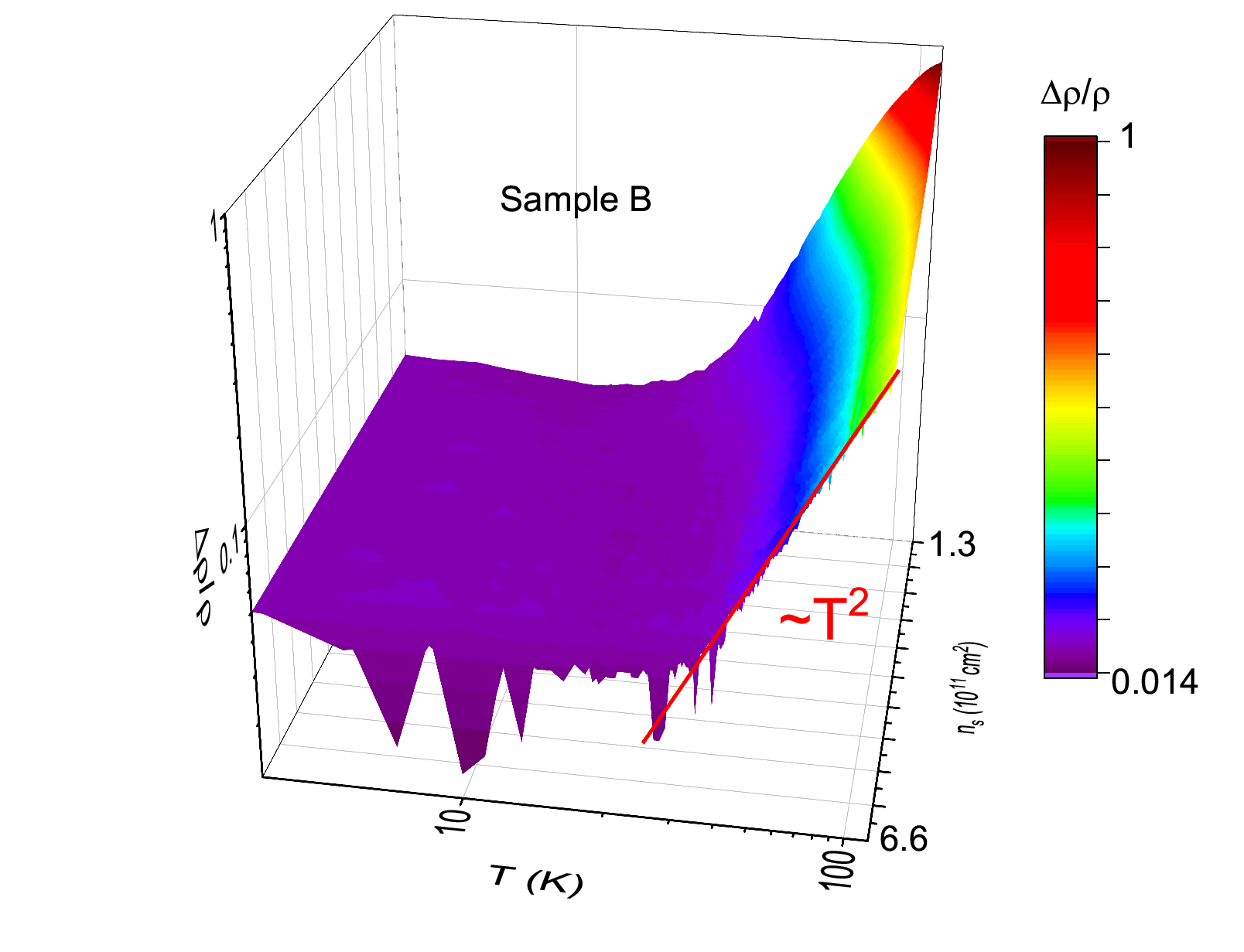}
\caption{ Excess of the resistivity as a function of the density and the temperature for sample B. Red line -$T^2$ dependence.}
\label{res2}
\end{figure}
Fig.~\ref{Landau} illustrates the evolution of longitudinal resistance ($R_{xx}$) as a function of magnetic field ($B$) and carrier density ($n_s$). The plot reveals a pattern of stripes corresponding to resistance maxima and minima in the $B$-$n_s$ plane for electron-like states. Notably, there is no direct correspondence between the experimental $R_{xx}(B, n_s)$ data and the Landau level (LL) spectrum, owing to the oscillatory behavior of the Fermi energy. While the Dirac LL spectrum exhibits a square root dependence on the magnetic field, the experimental $R_{xx}(B, n_s)$ diagram demonstrates a linear $n_s$ versus $B$ relationship. The slopes of these stripes are determined by the LL filling factors ($\nu$) as $\frac{dB}{dn} = \frac{\nu e}{h}$. Importantly, the filling factors extracted from the stripe slopes align with those determined from the Hall resistance measurements. Since the Landau level (LL) fan diagram displays a single set of lines without any crossings with LL levels from the second subband within the density range $0 < n_s < 8\times 10^{11} cm^{-2}$, this confirms that the electron energies are confined below the second quantized subband. Consequently, our interpretation of the data presented in the main text is fully justified.

\subsection{Temperature dependence of the resistivity}
In the main text, we present our experimental results on resistance measurements as a function of temperature and density, as shown in Figs.~\ref{sample}a and b. While this representation is effective, alternative visualizations, such as 3D color plots, can offer additional insights. A 3D plot enables the simultaneous representation of three variables (e.g., resistance, density, and temperature), making it easier to identify trends or interactions that may be challenging to discern in two-dimensional plots.

Figs.~\ref{res1} and Fig.~\ref{res2} illustrate the excess resistivity recalculated from the experimental dataset presented in Figure 1 of the main text. These plots reveal that the resistivity exhibits a clear $T^2$ dependence across a wide range of temperatures and densities, highlighting the robustness of this behavior in the system.
\subsection{Logarithmic corrections to the resistivity at temperatures below 20 K.}

Weak localization (or antilocalization) has been extensively studied in \cite{kozlov}, where a detailed comparison with theoretical predictions revealed excellent agreement. It is well known that, in the case of a disordered two-dimensional (2D) metal, the quantum correction to the conductivity can be expressed as the sum of two contributions, originating from weak localization and electron–electron interactions:
\begin{equation}
    \delta \sigma = \delta \sigma_{\text{loc}} + \delta \sigma_{\text{int}}.
\end{equation}

The weak localization correction contributes as follows:
\begin{equation}
    \delta \sigma_{\text{loc}} = \alpha_{\text{loc}} \frac{e^2}{h \pi} \ln \left( \frac{k_B T}{T_0} \right),
\end{equation}
where the coefficient $\alpha_{\text{loc}}$ equals $1$ in the case of weak localization and $-1/2$ in the case of antilocalization.

In the diffusion regime ($T \tau \gg 1$), the correction to the conductivity due to electron–electron interactions can be expressed in a simplified form as follows:
\begin{equation}
    \delta \sigma_{\text{int}} = \alpha_{\text{int}} \frac{e^2}{h \pi} \left( k_B T \tau /h\right),
\end{equation}
where $\tau$ momentum relaxation time and $\alpha_{\text{int}}$ interaction coefficient. 
It was observed that resistance decreases logarithmically with increasing temperature, following the relationship $\Delta R\sim \Delta \sigma R^2 \sim ln(T)$. This behavior arises from the effects of localization and disorder corrections due to electron-electron (e-e) interactions. The logarithmic temperature correction is determined by the dominant role of interaction effects.
An decrease in resistivity of approximately $5–7\%$ was identified, which could account for the observed rise in resistivity within the 5–15 K temperature range. We propose that this behavior fully explains the deviations from the expected $T^2$ dependence in this temperature interval. At higher temperatures, the contribution from e-e collisions becomes the dominant mechanism.

\subsection{Comparison with the Hydrodynamic Regime.}
\begin{figure}
\label{tau}
\includegraphics[width=9cm]{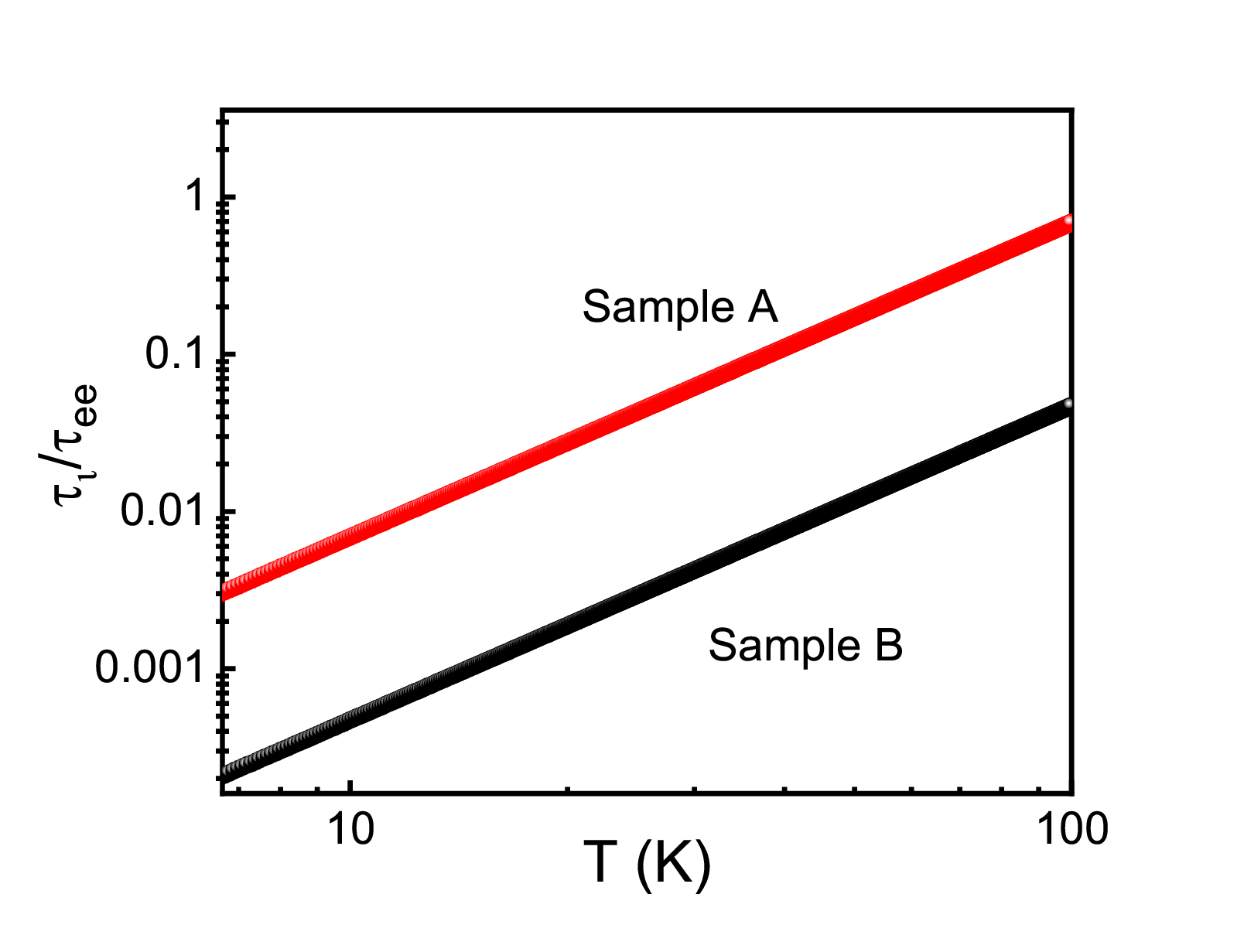}
\caption{The ratio $\tau_i / \tau_{ee}$ as a function of temperature  for sample A ($E_F=62.5 meV$) and B ($E_F=92.5 meV$) . }
\end{figure}
As discussed in the main text, electron-electron (e-e) collisions can contribute to the resistivity within the framework of theoretical models that consider the hydrodynamic regime. Two primary scenarios are typically analyzed:

The first scenario \cite{gurzhi} involves electron transport in narrow channels, driven by the Poiseuille flow of electrons. This phenomenon requires specific conditions: $l_{ee} < W < l$, where $l = v_F \tau$ is the electron mean free path associated with the momentum relaxation time $\tau$, $v_F$ is the Fermi velocity, $W$ is the channel width, and $l_{2,ee} = v_F \tau_{2,ee}$ is the mean free path for shear viscosity relaxation.  The subscript ``2'' indicates that the viscosity coefficient is determined by the relaxation of the second harmonic in the distribution function ( in the high-temperature limit $\tau_{2,ee} \approx \tau_{ee}$). However, this mechanism is not applicable to our macroscopic samples because their width satisfies the condition $W \gg l$.

It is also important to emphasize that in narrow channels, e-e interactions are the dominant mechanism, leading to an inverse temperature dependence resistivity $\sim T^{-2}$, known as the Gurzhi effect. It is because resistivity is instead governed by the Navier-Stokes equation rather than the Boltzmann equation \cite{gurzhi}. As a result, it follows the relation $\rho \sim \nu \sim T^{-2}$, where $\nu =\frac{1}{4}v_{F}^2 \tau_{ee}$ is the shear viscosity ($v_{F}$ is the Fermi velocity). This behavior occurs regardless of the energy spectrum. By contrast, for macroscopic samples with a parabolic spectrum, e-e interactions do not contribute to resistivity.

Only a limited number of system satisfy the conditions $l_{ee} < l$ or $\tau_{ee} < \tau$ within the low-temperature range, as these conditions typically require higher temperatures. Materials such as graphene and GaAs structures are among the few that meet these criteria.

In our case, the condition $\tau_{ee} > \tau$ or $\tau_i$ is necessary to apply Coulomb scattering within the framework of perturbation theory, as discussed in the main text. Figure~\ref{tau} illustrates the ratio $\tau_i / \tau_{ee}$ as a function of temperature. The figure clearly demonstrates that this ratio remains below $1$ throughout the entire temperature range for both samples. Therefore, our system does not satisfy the hydrodynamic conditions within this range of temperatures and densities.

The second scenario, proposed in theoretical studies \cite{spivak}, considers the emergence of Poiseuille flow in strongly inhomogeneous samples, even when the sample width is much larger than other characteristic lengths. This case is analogous to the first, as the model predicts a similar dependence of resistivity on shear viscosity and an inverse temperature dependence, $\rho \sim \nu \sim T^{-2}$, as in the Gurzhi effect.

However, both scenarios contradict our experimental observations, which exhibit a $T^2$ dependence. Therefore, these mechanisms cannot be considered viable explanations for the observed behavior.


\begin{thebibliography}{999}

\bibitem{haas} W. J. de Haas, J. de Boer, G. J. van dën Berg, The electrical resistance of gold, copper and lead at low temperatures, Physica, {\bf 1}, 609 (1934).

\bibitem{pinski}
F. J. Pinski, P. B. Allen, and W. H. Butler, Calculated electrical and thermal resistivities of Nb and Pd, Phys. Rev. B {\bf 23}, 5080 (1981).
\bibitem{behnia}
 Kamran Behnia, On the Origin and the Amplitude of T-Square Resistivity in
 Fermi Liquids,  Ann. Phys. (Berlin), {\bf 534}, 2100588 (2022).
\bibitem{pal}
H. K. Pal, V. I. Yudson, and D. L. Maslov, Reistivity of non
Galilean invariant term Fermi and non Fermi liquids, Lith. J. Phys. {\bf 52}, 142 (2012).
\bibitem{olshanetsky}
E. B. Olshanetsky, Z. D. Kvon, M. V. Entin, L. I. Magarill, N. N. Mikhailov, and
S. A. Dvoretsky. Scattering processes in a two-dimensional semimetal, JETP Lett.  {\bf 89}, 290 (2009).
\bibitem{entin}
M. V. Entin, L. I. Magarill, E. B. Olshanetsky, Z. D. Kvon, N. N. Mikhailov, and
S. A. Dvoretsky, The effect of electron-hole scattering on transport properties of a 2D semimetal in the HgTe quantum well. JETP,  {\bf 117}, 933 (2013).

\bibitem{hwang}
E. H. Hwang and S. Das Sarma, Temperature dependent resistivity of spin-split subbands in GaAs two-dimensional hole systems, Phys. Rev. B {\bf 67}, 115316 (2003).

\bibitem{nagaev1}
K. E. Nagaev, and A. A. Manoshin, Electron-electron scattering and transport properties of spin-orbit coupled electron gas,
Phys. Rev. B {\bf 102}, 155411 (2020).

\bibitem{nagaev2}
K. E. Nagaev, Electron-electron scattering and conductivity of disordered systems with a Galilean-invariant spectrum, Phys. Rev. B {\bf 106}, 085411 (2022).
\bibitem{gusev}
G. M. Gusev,  A. D. Levin, E. B. Olshanetsky, Z. D. Kvon, V. M. Kovalev, M. V. Entin,  and N. N. Mikhailov, Interaction-dominated transport in two-dimensional conductors: From degenerate to partially degenerate regime, Phys.Rev.B, 2024, {\bf 109}, 035302 (2024).
\bibitem{levin}
A. D. Levin, G. M. Gusev, F. G. G. Hernandez, E. B. Olshanetsky, V. M. Kovalev, M. V. Entin,  and N. N. Mikhailov, Interaction-controlled transport in a two-dimensional massless-massive Dirac system: Transition from degenerate to nondegenerate regimes, Phys.Rev.Research, {\bf 6}, 023121 (2024).

\bibitem{gurzhi}
R. N. Gurzhi, Minimum of Resistance in Impurity-free Conductors, Sov. Phys. JETP {\bf 44}, 771 (1963);
Sov. Phys. Usp. {\bf 11}, 255 (1968).

\bibitem{polini}
Marco Polini, and Andre K. Geim, Viscous electron fluids, Physics Today {\bf 73}, 6, 28 (2020).

\bibitem{narozhny}
Boris N. Narozhny, Hydrodynamic approach to two-dimensional electron systems, La Rivista del Nuovo Cimento  {\bf 45},  661 (2022).

\bibitem{dejong}
M. J. M. de Jong and L. W. Molenkamp, Hydrodynamic electron flow in high-mobility wires, Phys. Rev. B {\bf 51}, 13389 (1995).
\bibitem{kumar}
R. Krishna Kumar, D. A. Bandurin, F. M. D. Pellegrino, Y. Cao, A. Principi, H. Guo, G. H. Auton, M. Ben Shalom, L. A. Ponomarenko, G. Falkovich, K. Watanabe, T. Taniguchi, I. V. Grigorieva, L. S. Levitov, M. Polini, and A. K. Geim, Superballistic flow of viscous electron fluid through graphene constrictions, Nat. Phys. {\bf 13}, 1182 (2017).
\bibitem{spivak}
A. V. Andreev, Steven A. Kivelson, and B. Spivak, Hydrodynamic Description of Transport in Strongly Correlated Electron Systems, Phys. Rev. Lett. {\bf 106}, 256804 (2011)
\bibitem{gusev1}
G. M. Gusev, A. D. Levin, E. V. Levinson, and A. K. Bakarov, Viscous electron flow in mesoscopic two-dimensional electron gas, AIP Adv. {\bf 8},
025318 (2018).

\bibitem{gusev2}
G. M. Gusev, A. D. Levin, E. V. Levinson, and A. K. Bakarov, Viscous transport and Hall viscosity in a two-dimensional electron system, Phys. Rev. B {\bf 98}, 161303(R) (2018).

\bibitem{gusev3}
G. M. Gusev, A. S. Jaroshevich, A. D. Levin, Z. D. Kvon  and A. K. Bakarov, Stokes flow around an obstacle in viscous two-dimensional electron liquid, Sci.Rep. {\bf 10}, 7860 (2020).

\bibitem{gusev4}
G. M. Gusev, A. S. Jaroshevich, A. D. Levin, Z. D. Kvon  and A. K. Bakarov, Viscous magnetotransport and Gurzhi effect in bilayer electron system, Phys. Rev. B {\bf 103}, 075303 (2021).

\bibitem{bolotin}
K. I. Bolotin, K. J. Sikes, J. Hone, H. L. Stormer, and P. Kim, Temperature-Dependent Transport in Suspended Graphene, Phys.Rev.Lett {\bf 101}, 096802 (2008).
\bibitem{wagner}
G. Wagner, D. X. Nguyen, S. H. Simon, Transport in bilayer graphene near charge 
neutrality: Which scattering mechanisms are important? Phys. Rev. Lett. {\bf 124}, 026601 (2020)
\bibitem{nam}
Y. Nam, D.-K. Ki, D. Soler-Delgado, A. F. Morpurgo, Electron-hole collision limited 
transport in charge-neutral bilayer graphene. Nat. Phys. {\bf 13}, 1207 (2017).
\bibitem{tan}
Cheng Tan, Derek Y. H. Ho, Lei Wang, Jia I. A. Li, Indra Yudhistira, Daniel A. Rhodes,
Takashi Taniguchi, Kenji Watanabe, Kenneth Shepard, Paul L. McEuen, Cory R. Dean,
Shaffique Adam, James Hone, Dissipation-enabled hydrodynamic conductivity in a tunable bandgap semiconductor,  Sci. Adv. {\bf 8}, 8481 (2022).
\bibitem{suppl} 
See Supplemental Material for a detailed solution of the Boltzmann equation related to electron-impurity and electron-electron scattering, as well as descriptions of the samples and measurement details. The Supplemental Material also contains Refs. [12, 17, 30]
\bibitem{Fink1} Woo-Ram Lee, Alexander M. Finkel'stein, Karen Michaeli, and Georg Schwiete, 
Role of electron-electron collisions for charge and heat transport at intermediate temperatures, Phys. Rev. Research 2, 013148 (2020).

\bibitem{Fink2} Woo-Ram Lee, Alexander M. Finkel'stein, and Georg Schwiete, 
Role of electron-electron collisions for magnetotransport at intermediate temperatures, 
Phys. Rev. B 102, 245117 (2020).

\bibitem{buttner}
B. Büttner, C. X. Liu, G. Tkachov, E. G. Novik, C. Brüne, H. Buhmann, E. M. Hankiewicz, P. Recher, B. Trauzettel, S. C. Zhang and L. W. Molenkamp, Single valley Dirac fermions in zero-gap HgTe quantum wells, Nature Physics, {\bf 7}, 418, (2011).
\bibitem{kozlov}
D. A. Kozlov, Z. D. Kvon, N. N. Mikhailov, and S. A. Dvoretskii, Weak localization of Dirac fermions in HgTe quantum wells, JETP Lett. {\bf 96}, 730 (2012).
 \bibitem{alekseev}
 P. S. Alekseev and A. P. Dmitriev,  Viscosity of two-dimensional electrons, Phys. Rev. B {\bf 102}, 241409(R), (2020)
 \bibitem{narozhny2}
 B. N. Narozhny and M. Schutt, Magnetohydrodynamics in
graphene: Shear and Hall viscosities, Phys. Rev. B {\bf 100}, 035125 (2019)

\bibitem{melezhik}
Ye. O. Melezhik, J. V. Gumenjuk-Sichevska, and F. F. Sizov,
Electron relaxation and mobility in the inverted band quantum well $CdTe/Hg_{1-x}Cd_{x}Te/CdTe$, Semicond. Phys., Quantum Electron. Optoelectron. {\bf 17}, 85 (2014).
\bibitem{efetov}
Alexandre Jaoui, Ipsita Das, Giorgio Di Battista, Jaime Díez-Mérida,Xiaobo Lu, Kenji Watanabe, Takashi Taniguchi, Hiroaki Ishizuka, Leonid Levitov  and Dmitri K. Efetov, Quantum critical behaviour in magic-angle twisted bilayer graphene, Nature Physics {\bf 18}, 633 (2022).
\bibitem{kis}   
Sajedeh Manzeli, Dmitry Ovchinnikov, Diego Pasquier, Oleg V. Yazyev and Andras Kis, 2D transition metal dichalcogenides, Nature Reviews Materials {\bf 2}, Article number: 17033 (2017). 
\bibitem{stemmer}
Luca Galletti, Timo Schumann, Omor F. Shoron, Manik Goyal, David A. Kealhofer, Honggyu Kim, and Susanne Stemmer, Two-dimensional Dirac fermions in thin films of $Cd_3As_2$, Phys. Rev. B \textbf{97}, 115132 (2018).
\bibitem{hasan}
M. Z. Hasan, C. L. Kane, Colloquium: Topological insulators, Rev. Mod. Phys., {\bf 82}, 3045 (2010).
\bibitem{kotov}
Valeri N. Kotov, Bruno Uchoa, Vitor M. Pereira, F. Guinea, and A. H. Castro Neto, Electron-Electron Interactions in Graphene: Current Status and Perspectives, Rev. Mod. Phys. {\bf 84} , 1067 (2012).
\bibitem{elias}
D. C. Elias, R. V. Gorbachev, A. S. Mayorov, S. V. Morozov, A. A. Zhukov, P. Blake, L. A. Ponomarenko, I. V. Grigorieva, K. S. Novoselov, F. Guinea  and A. K. Geim, Dirac cones reshaped by interaction effects in suspended graphene, Nature Physics {\bf 7},  701 (2011).
\bibitem{xin}
Na Xin, James Lourembam, Piranavan Kumaravadivel, A. E. Kazantsev, Zefei Wu, 
Ciaran Mullan, Julien Barrier, Alexandra A. Geim, I. V. Grigorieva, A. Mishchenko, 
A. Principi, V. I. Fal’ko1, L. A. Ponomarenko, A. K. Geim1, and Alexey I. Berdyugin,
Giant magnetoresistance of Dirac plasma in high-mobility graphene, Nature {\bf 616}, 270–274 (2023).
\end{thebibliography}
\end{document}